\documentclass[twocolumn,twoside,slac_two]{revtex4}
\usepackage{graphicx}
\usepackage{fancyhdr}
\usepackage{gensymb}
\usepackage{lineno}
\newcommand{\Fermi}{{\it Fermi }}
\newcommand{\fermi}{{\it Fermi}}
\newcommand{\FermiL}{{\it Fermi}-LAT }
\newcommand{\fermiL}{{\it Fermi}-LAT}
\newcommand{\Off}{{\it Off }}

\newcommand{\On}{{\it On }}

\def\apj{ApJ }				% Astrophysical Journal
\def\apjl{ApJ }				% Astrophysical Journal, Letters
\def\apjs{ApJS }				 % Astrophysical Journal, Supplement
\def\aap{A\&A } 				% Astronomy and Astrophysics
\def\bams{\ref@jnl{Bull.~Am.~Meteorol.~Soc.}} % Bulletin of the American Meteoological Society
\def\bssa{\ref@jnl{Bull.~Seismol.~Soc.~Am.}} % Bulletin of the Seismological Society of Americ
\def\dsr{\ref@jnl{Deep~Sea~Res.}}	% Deep Sea Research
\def\eos{\ref@jnl{Eos~Trans.~AGU}}	% Eos (Transactions of the AGU)
\def\epsl{\ref@jnl{Earth~Planet.~Sci.~Lett.}}	% Earth and Planetary Sciences Letters
\def\gca{\ref@jnl{Geochim.~Cosmochim.~Acta}}	% Geochimica & Cosmochimica Acta
\def\gjras{\ref@jnl{Geophys.~J.~R.~Astron.~Soc.}} % Geophysical Journal of the RAS
\def\grl{\ref@jnl{Geophys.~Res.~Lett.}}	% Geophysics Research Letters
\def\gsab{\ref@jnl{Geol.~Soc.~Am.~Bull.}}	% Bulletin of the GSA
\def\jatp{\ref@jnl{J.~Atmos.~Terr.~Phys.}}	% Journal of Atmospheric and Terrestrial Physics
\def\jgr{\ref@jnl{J.~Geophys.~Res.}}	% Journal of Geophysical Research
\def\jpo{\ref@jnl{J.~Phys.~Oceanogr.}}	% Journal of Physical Oceanography
\def\mnras{MNRAS}			% Monthly Notices of the RAS
\def\mwr{\ref@jnl{Mon.~Weather~Rev.}}	% Monthly Weather Review
\def\nat{Nature }
\def\prd{Phys.~Rev.~D }		% Physical Review D
\def\qjrms{\ref@jnl{Q.~J.~R.~Meteorol.~Soc.}}	% Quarterly Journal of the RMS
\def\rg{\ref@jnl{Rev.~Geophys.}}	% Review of Geophysics
\def\rs{\ref@jnl{Radio~Sci.}}		% Radio Science
\def\usgsof{\ref@jnl{U.S.~Geol.~Surv. Open File~Rep.}}	% USGS Open File Report
\def\usgspp{\ref@jnl{U.S.~Geol.~Surv.~Prof.~Pap.}}	% USGS Professional Papers
\begin{document}

\title{Pulsar Emission above the Spectral Break - A Stacked Approach }


\author{A. McCann}
\affiliation{EFI \& KICP at The University of Chicago, Chicago, IL, 60637, USA}
%\affiliation{Colalborative University/Institute, City, State, Postal Code Country}

\begin{abstract}
NASA's \Fermi space telescope has provided us with a bountiful new
population of gamma-ray sources following its discovery of over 150
new gamma-ray pulsars. One common feature exhibited by all of these
pulsars is the form of their spectral energy distribution, which can
be described by a power law followed by a spectral break occurring
between $\sim$1 and $\sim$8 GeV. The common wisdom is that the break
is followed by an exponential cutoff driven by
radiation-reaction-limited curvature emission. The discovery of pulsed
gamma rays from the Crab pulsar, the only pulsar so far detected at
very high energies (E$>$100~GeV), contradicts this ``cutoff''
picture. Here we present a new stacked analysis with an average of 4.2
years of data on 115 pulsars published in the 2nd \FermiL catalog of
pulsars. This analysis is sensitive to low-level $\sim$100 GeV
emission which cannot be resolved in individual pulsars but can be
detected from an ensemble.
%We also report on pulsar observations with
%the VERITAS telescope, including new upper limits on pulsed emission
%from Geminga.
\end{abstract}

\maketitle

\thispagestyle{fancy}


\section{Introduction}
\begin{figure*}
\centering
\includegraphics[width=0.9\textwidth]{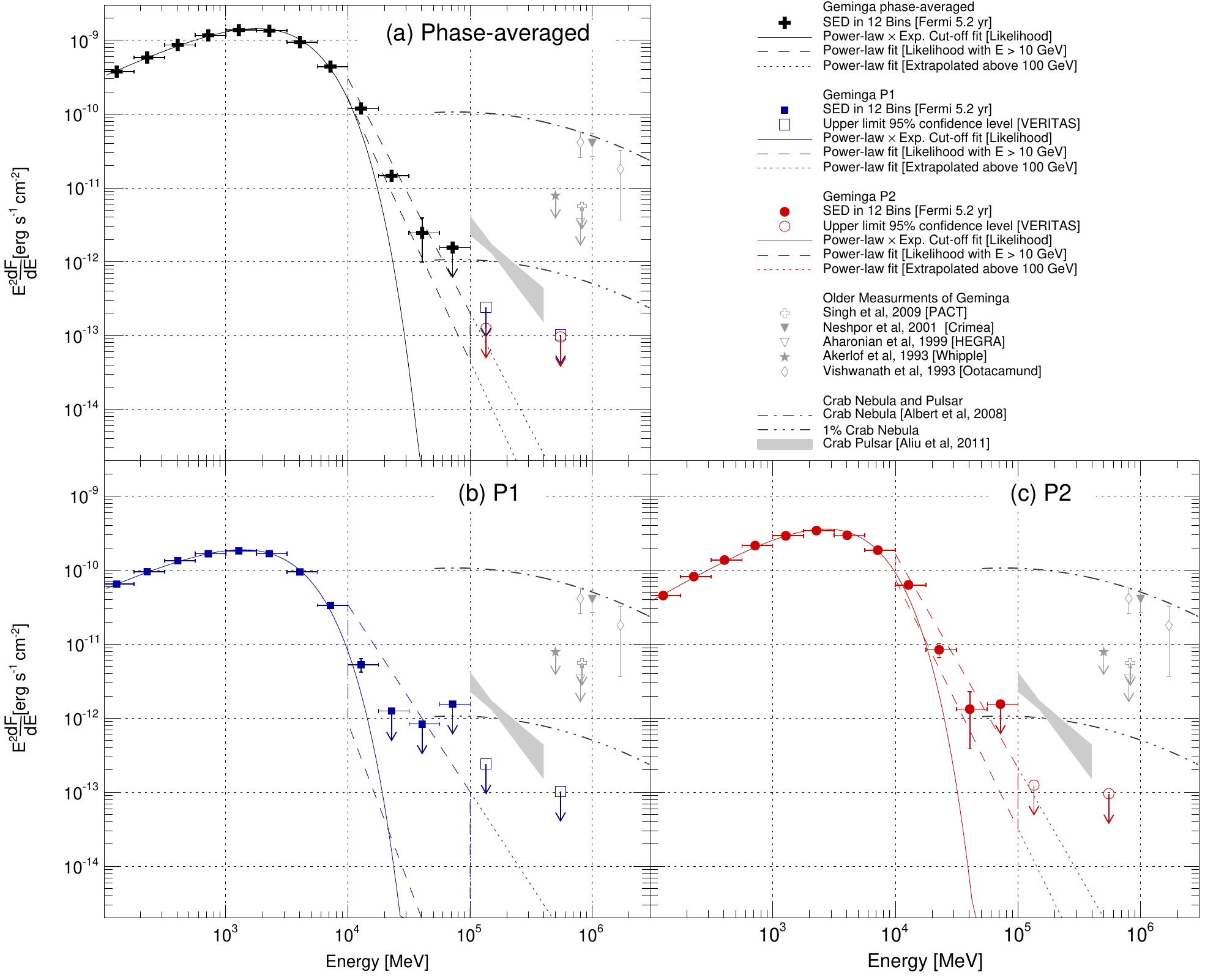}[ht]
\caption{SEDs data points and flux upper limits for the Geminga
  pulsar. Measurements of the Crab Nebula and pulsar are plotted for
  comparison. It is clear, even in the phase-resolved analysis, that
  the SED falls slower than an exponential and appears more consistent
  with a simple power-law. Figure taken from \cite{Aliu2014Gem}. }
\end{figure*}
One common feature exhibited by all known gamma-ray pulsars is the
form of their spectral energy distribution (SED) which can be
described by a power-law followed by a spectral break occurring
between 1 and 8~GeV \citep{Abdo2013ApJS}. The unanimity of the break
energy across the entire \FermiL pulsar sample is suggestive that the
sites of acceleration and processes of gamma-ray emission are common
across different pulsar types and that they are not strongly dependent
on the pulsar spin or energetics. Further, it has been shown that
across the \FermiL pulsar sample the spectral-break energy is weakly
correlated with the magnetic-field strength at the light cylinder
\citep{Abdo2010ApJS, Abdo2013ApJS}. Such behavior is expected in
models where emission is produced by curvature radiation (CR)
occurring at the radiation-reaction limit in the outer magnetosphere
\citep{Harding2008ApJ,Abdo2010ApJS}. This has become the most favored
general description of gamma-ray emission from pulsars in the \FermiL
era. In these models one expects that the SED will fall off
exponentially above the break energy. There is, however, compelling
evidence suggesting that CR occurring in the outer
magnetosphere is not a complete description of pulsar emission at, and
above, the GeV SED break:
\begin{enumerate}
\item The discovery of power-law-type emission from the Crab pulsar at
  energies exceeding 100~GeV\footnote{At this symposium the MAGIC
    collaboration presented evidence indicating that the power-law
    spectrum of the Crab pulsar may extend to TeV energies. See
    \url{http://fermi.gsfc.nasa.gov/science/mtgs/symposia/2014/abstracts/185}}
  cannot be easily explained by curvature emission from the outer
  magnetosphere \citep{Aliu2011Sci,Lyutikov2012ApJb} unless the radius
  of curvature of the magnetic field line is larger than the radius of
  the light cylinder \citep{Bednarek2012}. Some recent models
  attribute the pulsed very-high-energy (VHE; $E>$100~GeV) emission
  from the Crab pulsar to inverse-Compton (IC) scattering originating
  in the outer magnetosphere
  \citep{Lyutikov2012ApJb,Du2012ApJ,Lyutikov2012ApJ} or to IC
  scattering from beyond the light cylinder
  \citep{Aharonian2012Natur,Petri2012MNRAS}.
\item The radiation-reaction limit of CR occurs when the acceleration
  gains achieved by an electron are equaled by radiation losses. The
  photon energy at which this occurs in the outer magnetosphere can be
  expressed in terms of the pulsar period, the surface magnetic field
  strength, the radius of curvature of the accelerated particle and
  an efficiency factor. \cite{Lyutikov2012ApJb} has shown that the
  break-energy values for several pulsars reported in the first
  \FermiL pulsar catalog are so high that they require the efficiency
  factor and radius of curvature to approach or even reach their
  maximal allowable values\footnote{In more realistic models the
    acceleration efficiency is expected to be a few percent to a few
    tens of percent \citep{Lyutikov2012ApJb}.}.
\item Recent studies of the Geminga pulsar with \FermiL and VERITAS
  (see Figure~1) show that the SED above the GeV break is compatible
  with a steep power law \citep{Lyutikov2012ApJ,Aliu2014Gem}, but no
  emission has been seen above 100~GeV. Similar conclusions can be
  drawn from an analysis of the Vela pulsar with \FermiL data from
  \cite{Leung2014}, who show that multi-zone or time-dependent
  emission models are needed to fit the slower-than-exponential fall
  of the SED above 10~GeV.
\end{enumerate}
The question of whether the Crab pulsar is unique, or whether
non-exponentially-suppressed gamma-ray spectra are common in gamma-ray
pulsars is of great importance. Beyond the modeling of pulsar
emission, questions concerning the emission spectra of pulsars have
significant implications for galactic dark matter searches, where
unassociated gamma-ray excesses can be interpreted as the remnants of
dark matter annihilation (e.g., \citealt{Abazajian2012PhRvD}). Since
pulsars are likely the main background for these searches,
categorizing the shape of pulsar spectra is a critical step towards
validating any indirect dark matter signal in the gamma-ray domain.
To search for non-exponentially-suppressed emission above 50~GeV, we
have performed a stacked analysis of gamma-ray pulsars which is
sensitive to emission which cannot be resolved in the \FermiL analysis
of individual objects, but can be detected if aggregated from an
ensemble. A stacked analysis which yields evidence of cumulative
emission above 50~GeV would prove that some population of gamma-ray
pulsars clearly exhibits non-exponentially-suppressed emission. This
would indicate that inverse-Compton or wind-zone emission is common in
gamma-ray pulsars and that pulsars contribute to the sub-TeV diffuse
emission of the galaxy.

\section{Analysis}
\begin{figure*}
\centering
\includegraphics[width=\textwidth]{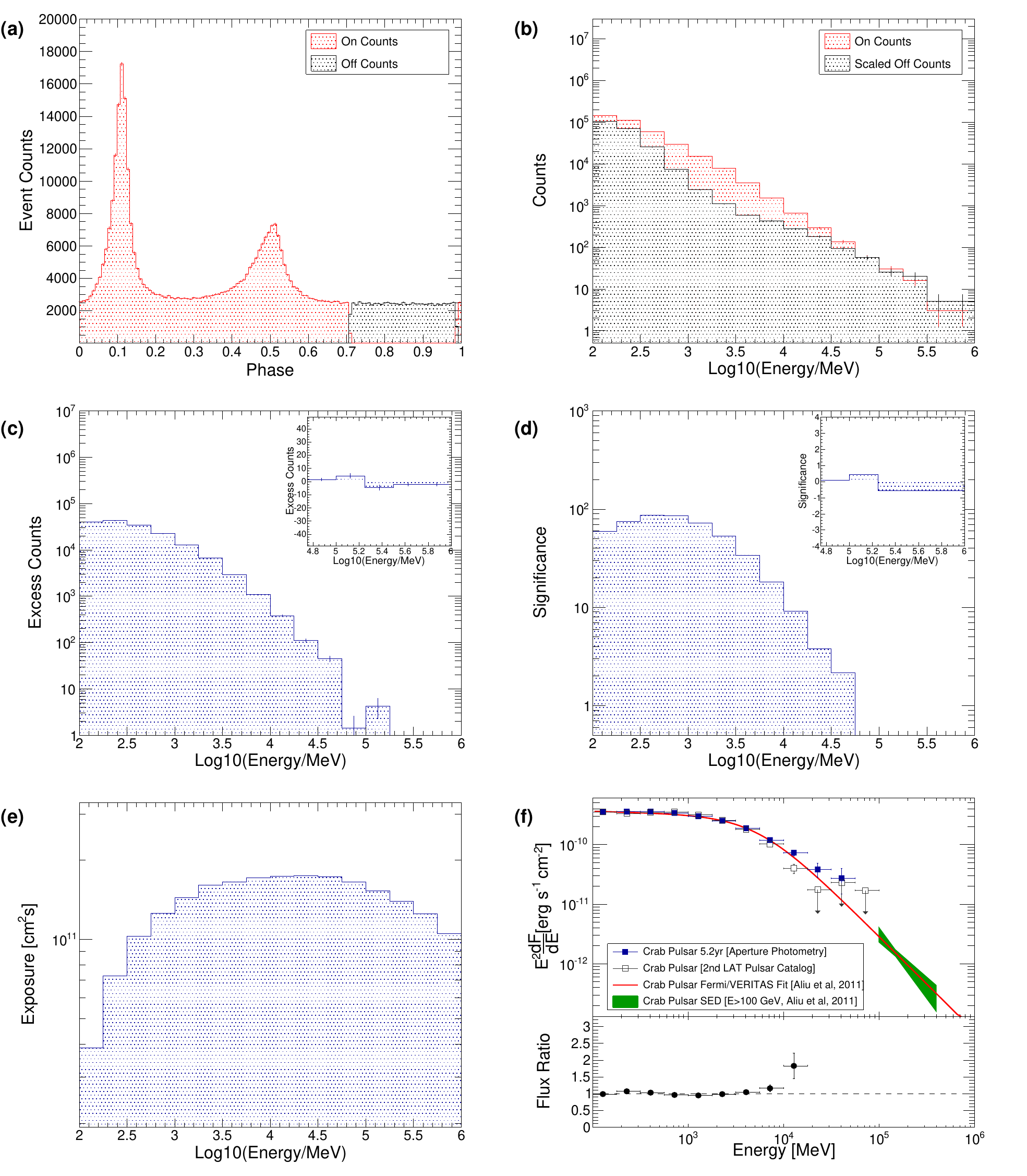}
\caption{Aperture photometry analysis steps for the Crab pulsar. Panel
  (a) plots the phase distribution (light curve) of the Crab pulsar
  from 5.2 years of \fermi-LAT observations. The \Off phase range,
  [0.71 $-$ 0.99], is defined in the 2nd \fermi-LAT catalog of
  gamma-ray pulsars (2PC). Panel (b) plots the distribution of photon
  energies for events which fell in the \On and \Off phase ranges. The
  \Off events have been scaled by $\alpha$ which is the ratio of the
  \On phase gate(s) size to the \Off gate(s) size . Panel (c) shows
  the energy distribution of the excess events and panel (d) shows the
  significance of the excess in each energy bin. Panel (e) shows the
  \fermi-LAT exposure for the ROI used in each energy bin determined
  from \texttt{gtexposure}. In panel (f) the Crab pulsar AP SED is
  plotted alongside the Crab pulsar SED determined from a likelihood
  fit done in the 2PC. A broken power-law fit to \FermiL and VERITAS
  data from \cite{Aliu2011Sci} is plotted, as well as the VERITAS
  $>$100~GeV bow-tie. Below the SED plotted in panel (f) is the ratio
  of the AP flux to the 2PC flux in each bin, showing the level of
  agreement between the AP method and the likelihood method. This
  figure is taken from \cite{arXiv1412.2422M}.} \label{JACpic2-f1}
\end{figure*}
\subsection{The Aperture Photometry Method}\label{sec:APmeth}
A maximum likelihood fitting procedure is typically employed when
performing spectral analysis of \fermi-LAT data. The \fermi-LAT data
can also be analyzed with an \textit{Aperture Photometry} (AP) method
where the raw event counts from a region of interest (ROI) are
combined with a measure of the instrument exposure (cm$^{2}$\,s) to
the region to determine the flux. This AP method is less sensitive and
less accurate than the likelihood fitting procedure but it ``provides
a model independent measure of the flux'' and it ``is less
computationally
demanding''\footnote{\url{http://fermi.gsfc.nasa.gov/ssc/data/analysis/scitools/aperture_photometry.html}}. We
demonstrate here that the AP method can be used to produce accurate
SEDs from multi-year pulsar data sets since an accurate determination
of the background rate can be measured from the ``\textit{Off}-pulse''
phase range. The analysis presented here proceeds as follows:
\begin{enumerate}
\item Over the 100~MeV to 1~TeV energy range, logarithmically-spaced
  energy binning with 4 bins per decade is chosen.
\item An ROI is chosen around each pulsar with an energy-dependent
  radius. The radius chosen is three times the 68\%
  point-spread-function (PSF) containment radius determined from a
  ``front-conversion'' Vela analysis by \cite{Ackermann2013ApJ}. In
  order to maintain sufficient statistics at high energies, the radius
  of the ROI was fixed to 0.45$\degree$ above 10~GeV.
\item The \FermiL analysis tools \texttt{gtselect}, \texttt{gtmktime},
  \texttt{gtbin} and \texttt{gtexposure} are then run over each pulsar
  ROI for all observations performed within the period of validity of
  the pulsar timing solution.
\item The photon event list is barycentered and phase-folded using the
  \texttt{Tempo2} package \citep{Hobbs2006MNRAS} with the \Fermi
  \texttt{Tempo2} plugin and the corresponding timing solution.
\item Within each energy bin, a cut on phase is applied and events
  which fall within the \Off phase region and those which fall outside
  this region - the \On phase region - are selected. The ratio of the
  size of the \On phase range to the size of the \Off phase range,
  defined as $\alpha$, is then used to scale the number of event
  counts in the \Off phase region (N$_{\rm off}$) to the number in the
  \On region (N$_{\rm on}$).
\item The number of excess pulsed events is then defined as N$_{\rm
  ex}$$=$N$_{\rm on}$$-$$\alpha$N$_{\rm off}$ and the flux is N$_{\rm ex}$
  divided by the exposure ($\mathcal{T}$) calculated in step 3 using
  \texttt{gtexposure}. The significance of the excess is calculated
  using Equation~17 from \cite{LiMa1983ApJ}.
\end{enumerate}
Following this procedure one can derive the energy distributions for
the \On and \Off phase regions, and the instrument exposure, for any
pulsar. These distributions and the derived AP SED are shown for the
Crab pulsar in Figure~2.

\subsection{Stacking The Pulsar Data Sets}
\FermiL has detected over 150 new gamma-ray
pulsars\footnote{\url{https://confluence.slac.stanford.edu/display/GLAMCOG/Public+List+of+LAT-Detected+Gamma-Ray+Pulsars}}
and the stacking performed in this work uses 115 pulsars listed in the
Second \FermiL Catalog of Gamma-ray Pulsars \citep{Abdo2013ApJS},
which shall be referred to as 2PC throughout\footnote{A total of 117
  pulsars are listed in the 2PC, however, the Crab pulsar was excluded
  from this analysis since we are investigating whether high-energy
  Crab-pulsar-like emission is seen in other pulsars. Further,
  PSRJ2215+5135 was also excluded from the study since no \Off phase
  region was listed for this source in the 2PC.}. The 115 pulsar
sample is composed of 39 millisecond pulsars and 76 ``young''
non-recycled pulsars with an average data set spanning
4.2~yr\footnote{The amount of data analyzed here depends entirely on
  the availability and validity of pulsar spin-down timing solutions
  used for phase-folding.}. The six AP analysis steps listed in
Section~\ref{sec:APmeth} were followed for each pulsar and using the
resulting values of N$_{\rm on}$, N$_{\rm off}$ and $\mathcal{T}$, it
is quite simple to determine the total excess,
\begin{equation}
{\rm Ex}_{\rm tot}^{i} = \sum_{j=1}^{\rm N}({\rm N}_{\rm on}^{j,i} - \alpha^{j}{\rm N}_{\rm off}^{j,i})
\end{equation}
the total exposure,
\begin{equation}
\mathcal{T}_{\rm tot}^{i} = \sum_{j=1}^{\rm N}\mathcal{T}^{j,i}
\end{equation}
and thus, the average flux,
\begin{equation}
{\rm Flux}_{\rm av}^{i} = \frac{{\rm Ex}_{\rm tot}^{i}}{\mathcal{T}_{\rm tot}^{i}}
\end{equation}
for N pulsars in a given energy bin, $i$. The significance of the
total excess is determined by the generalized version of Equation~17
from \cite{LiMa1983ApJ} (see \citealt{Aharonian2004A&A}). In cases
where the significance is less than 2$\sigma$, the method of
\cite{Helene1983} is used to derive the 95\% confidence-level upper
limit on the total excess, which is in turn used to compute a flux
upper limit.

\section{Results}
\begin{figure*}[t]
\centering
\includegraphics[width=\textwidth]{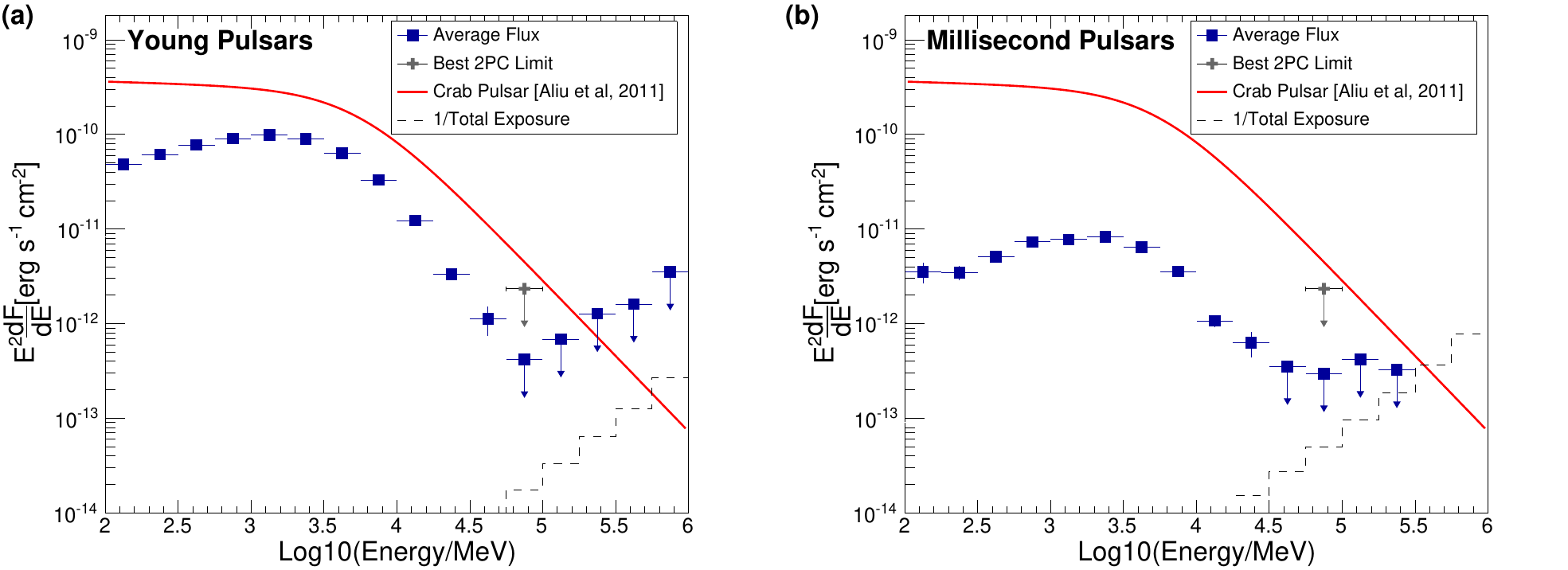}
\caption{Panel (a) shows the average flux (square markers) from 76
  young pulsars determined by dividing the total excess by the total
  exposure (see Equations~1$-$3).  The dashed-line histogram shows one
  over the total exposure, indicating the flux which would correspond
  to a single excess photon. This is the minimum possible flux which
  could be measured given the total exposure. The gray cross shows the
  most constraining limit on emission from a single pulsar in the
  56.2$-$100~GeV range presented in the 2PC. The 2PC presented no
  limits at higher energies. The broken power-law fit to the Crab
  pulsar data from \cite{Aliu2011Sci} is plotted for scale. Panel (b)
  plots the same quantities for the stacked analysis of 39 millisecond
  pulsars. This figure is adapted from figures presented in
  \cite{arXiv1412.2422M}.} \label{JACpic2-f1}
\end{figure*}
The stacking analysis results for the young pulsar and millisecond
pulsar ensembles are shown in Figure~3. No significant excesses are
seen in these analyses at energies above 50~GeV. Upper limits on the
average flux, determined at the 95\% confidence-level, are listed in
Table~\ref{tab:Stacksummary} for three energy bins above
50~GeV. Limits are also presented in units of the Crab pulsar where
the broken power-law fit to the \FermiL and VERITAS data presented in
\cite{Aliu2011Sci} defines a Crab pulsar unit. In addition to these
analyses, we stacked sub-samples of the data where each sub-sample was
composed of the 10 pulsars with the largest value of a given
parameter. Sub-sample selections based on gamma-ray luminosity,
spin-down power, spin-down power over distance squared, gamma-ray
photon flux and non-thermal X-ray energy flux were
investigated\footnote{The Crab pulsar was excluded from all of these
  sub-sample stacking analyses. The parameter values listed in the 2PC
  catalog were used in all cases.}. No significant excesses were
observed above 50~GeV in any of these sub-sample stacking analyses.

\begin{figure}
\centering 
\includegraphics[width=0.49\textwidth]{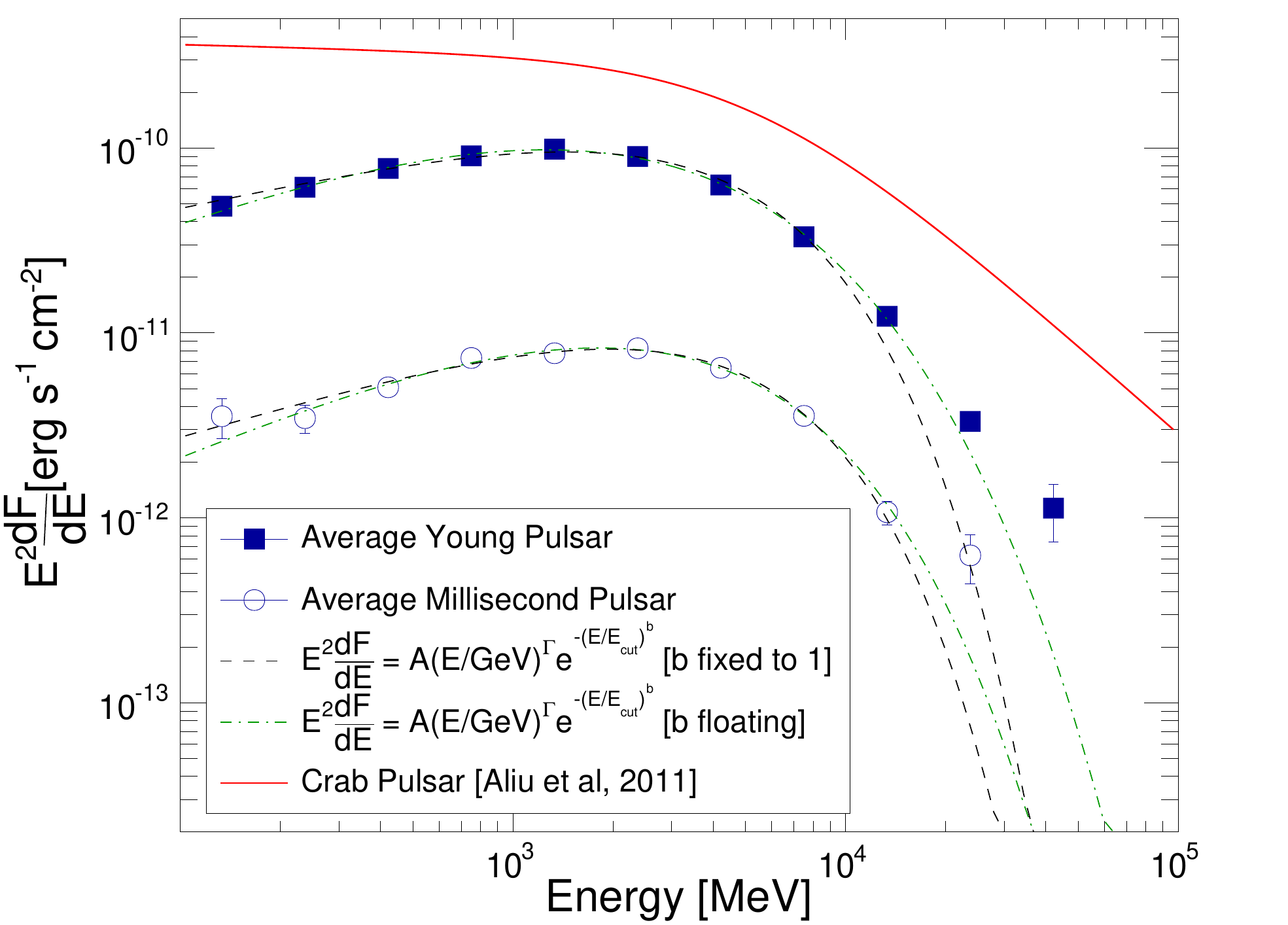}
\caption{\small The average SEDs derived from the stacking of the 76
  young pulsars and 39 millisecond pulsars. The SEDs are each fit with
  a power law times a super-exponential cutoff keeping $b$ both fixed
  to unity and allowing it to float. For the pure exponential cutoff
  case ($b=1$) the best fit $\Gamma$ value is 0.54$\pm$0.05 for the
  millisecond pulsars and 0.41$\pm$0.01 for the young pulsars while
  the best fit $E_{\rm cut}$ values are 3.60$\pm$0.21~GeV and
  3.54$\pm$0.04~GeV, respectively. Allowing $b$ to float we find that
  sub-exponential forms ($b<1$) are preferred, with the best-fit $b$
  value of 0.59$\pm$0.02 for the young pulsars and 0.7$\pm$0.15 for
  the millisecond pulsars. The broken power-law fit to the Crab pulsar
  data from \cite{Aliu2011Sci} is plotted for scale. Note that only
  statistical uncertainties on the SED data points were used during
  the fitting and thus the uncertainty on the best-fit parameter
  values are likely underestimated. This figure is taken from
  \cite{arXiv1412.2422M}.}
\label{fig:YPVMSP}
\end{figure}

The shape of the average young pulsar and average millisecond pulsar
SEDs were categorized by fitting a power law times a super-exponential
cutoff function
\begin{linenomath}
\begin{equation}
E^{2}\frac{dF}{dE} = A {\left(\frac{E}{\rm 1~GeV}\right)}^{\Gamma}e^{-\left(\frac{E}{E_{\rm cut}}\right)^{b}}
\end{equation}
\end{linenomath}
to the SED data. These fits are presented in
Figure~\ref{fig:YPVMSP}. Fixing $b=1$ reduces Equation~4 to a power
law times an exponential cutoff function and, as expected, this
functional form does not reproduce the sub-exponential fall of the SED
above the break. However it can be used to measure the average
flux-weighted value of the spectral index ($\Gamma$) and cutoff
($E_{\rm cut}$) parameters \citep{Abdo2013ApJS}. It is clear from
Figure~\ref{fig:YPVMSP} that the average SEDs have qualitatively the
same shape, with the average flux from the 39 millisecond pulsars
about an order of magnitude lower than the average flux from the 76
young pulsars. The spectral parameters derived from fitting the two
ensembles are remarkably similar and are presented in the caption of
Figure~\ref{fig:YPVMSP}.
\begin{table*}
\centering
\begin{tabular}{ccc|cc|cc|cc}\hline
      & &                          & \multicolumn{2}{|c|}{All}       & \multicolumn{2}{|c|}{Young Pulsars} & \multicolumn{2}{|c}{Millisecond Pulsars}   \\\hline
\multicolumn{3}{c|}{Energy Range} & Flux Limit       & Flux Limit   & Flux Limit      & Flux Limit    & Flux Limit       & Flux Limit   \\
\multicolumn{3}{c|}{[GeV]}        & [$\times10^{-12}$ & [Crab pulsar & [$\times10^{-12}$ & [Crab pulsar & [$\times10^{-12}$ & [Crab pulsar  \\
\multicolumn{3}{c|}{}             & cm$^{-2}$s$^{-1}$] &  units]      & cm$^{-2}$s$^{-1}$] &  units]      & cm$^{-2}$s$^{-1}$] &  units]      \\\hline
 56.2 &---& 100                        &  1.57            & 0.07         &  2.03            & 0.09         &  1.44            & 0.07         \\
  100 &---& 177                        &  1.52            & 0.31         &  1.88            & 0.38         &  1.14            & 0.23         \\ 
  177 &---& 316                        &  1.34            & 1.21         &  1.96            & 1.76         &  0.50            & 0.45         \\\hline
\end{tabular}
\caption{\small Limits at the 95\% confidence level on the average
  flux from stacked ensembles of gamma-ray pulsars. The limit values
  presented in Crab pulsar units assume the broken power-law fit to
  the Crab pulsar data from \cite{Aliu2011Sci} is a Crab pulsar flux
  unit.}
\label{tab:Stacksummary}
\end{table*}

\section{Discussion and Conclusion}
Following a stacked analysis of 115 gamma-ray pulsars, with an average
exposure of $\sim$4.2~yr per pulsar, we find no evidence of cumulative
emission above 50~GeV. Stacked searches exclusive to the young
pulsars, the millisecond pulsars, and several other promising
sub-samples also return no significant excesses above 50~GeV. Any
average emission present in the entire pulsar sample is limited to be
below $\sim$7\% of the Crab pulsar in the 56-100~GeV band. The average
flux limits presented in Table~\ref{tab:Stacksummary} are roughly 3
times lower than the best flux limits achieved in dedicated individual
pulsar analyses done in the 2PC in the 56-100~GeV band.

One should note that a limit on the average flux from 115 pulsars at
7\% of the Crab pulsar level is consistent with, for example, a
scenario in which all 115 pulsars emit at 7\% of the Crab pulsar
level. It is also consistent with a scenario in which 8 pulsars emit
at 100\% the level of the Crab pulsar and the remaining 107 pulsars
have zero emission. Therefore this analysis does not exclude the
possibility of finding several pulsars which are as bright as the Crab
pulsar above 50 GeV, or several dozen which are ten times dimmer. It
does, however, constrain the average flux from the ensemble, and
therefore for every individual pulsar detected above this flux limit,
the average emission from the remaining pulsars is constrained to be
further below the limit.

In the 100~MeV to $\sim$50~GeV energy range we find that the average
SEDs returned from the young pulsar and millisecond pulsar stacking
analyses are very similar in shape and are generally compatible with a
power law times a sub-exponential cutoff. \cite{Abdo2010ApJ} and
\cite{Celik2011AIPC} have shown that a sub-exponential cutoff function
approximates a superposition of exponential cutoffs, thus the
appearance of a sub-exponential cutoff in the ensemble SED is to be
expected within a curvature radiation model. We note, however, that
the highest energy spectral point is higher than the best fit
sub-exponential cutoff function at the $\sim$2.4$\sigma$ level in both
the young pulsar and millisecond pulsar cases. This cannot be taken as
strong evidence for a non-exponentially-suppressed pulsar emission
component aggregating in the stacked analysis, however, the available
data cannot rule it out beyond the level of the limits shown in
Figure~3 and Table~\ref{tab:Stacksummary}.

Beyond this work, improvements can be made using the forthcoming
\FermiL pass-8 data release which will improve the \fermi-LAT
acceptance by $\sim$25\% at 100~GeV
\citep{Atwood2013arXiv}. Improvements to this stacking analysis can
also be made by employing a likelihood framework to stack the sources
(see \citealt{Ackermann2011PhRvl} for example), rather than the simple
\On minus \Off procedure described here. The flux sensitivity of any
stacking analysis will, however, ultimately be bounded by the exposure
of the \fermiL. A future stacking analysis which doubles both the
number of pulsars and the duration of observation used will increase
the exposure by a factor of 4, indicating that future stacking
analyses which do not yield detections may improve on the limits
presented here by perhaps one or two orders of magnitude.

A more detailed account of the stacking analysis methods and
results of this study can be found in \cite{arXiv1412.2422M}.

\bigskip % extra skip inserted
\begin{acknowledgments}
This research is supported in part by the Kavli Institute for
Cosmological Physics at the University of Chicago through grant NSF
PHY-1125897 and an endowment from the Kavli Foundation and its founder
Fred Kavli. I am grateful to the VERITAS collaboration for their
support. In particular, I am very grateful to Pat Moriarty, David
Hanna, Nepomuk Otte, Benjamin Zitzer, Jeremy Perkins, Scott Wakely,
Nahee Park, Jeffery Grube and Christopher Bochenek for very helpful
discussions on aspects of the analysis presented here.

\end{acknowledgments}

\bigskip % extra skip inserted

\end{document}